# Cooperative behavior of quantum dipole emitters coupled to a zero-index nanoscale waveguide

*Ruzan Sokhoyan and Harry A. Atwater*

Thomas J. Watson Laboratories of Applied Physics, California Institute of Technology, Pasadena, California 91125, USA

ABSTRACT: We study cooperative behavior of quantum dipole emitters coupled to a rectangular waveguide with dielectric core and silver cladding. We investigate cooperative emission and inter-emitter entanglement generation phenomena for emitters whose resonant frequencies are near the frequency cutoff of the waveguide, where the waveguide effectively behaves as zero-index metamaterial. We show that coupling emitters to a zero-index waveguide allows one to relax the constraint on precision positioning of emitters for observing inter-emitter entanglement generation and extend the spatial scale at which the superradiance can be observed.

KEYWORDS: entanglement, superradiance, effective zero index, quantum optics, plasmonics

Coupling quantum emitters to a common reservoir, results in an effective interaction between them. The characteristics of the interaction are defined both by properties of the reservoir and the emitters. For example, for the emitters in the free space inter-emitter interactions diminish drastically when inter-emitter distance exceeds half of the resonant wavelength [1]. Coupling emitters to an environment with carefully designed local density of optical states (LDOS), such as a metamaterial or a waveguide, could substantially increase their interaction range. For example entanglement generation between two quantum emitters, placed ten free space wavelengths apart from each other, by using left-handed materials has been discussed [2].

Deep subwavelength field localization available with surface plasmons (SPs) opens up the prospect of miniaturization and scalability, beyond the limits of conventional photonic systems, that can be used for quantum optics applications. The possibility of utilizing SPs for enhancing the coupling between quantum emitters has been addressed by different authors [3-7]. For example, resonance energy transfer and superradiance assisted by plasmonic waveguides has been recently studied, and the coupling of emitters at separations much larger than involved optical wavelength has been demonstrated [3]. Gonzalez-Tudela and co-workers have investigated entanglement dynamics of quantum emitters coupled to a one-dimensional plasmonic waveguide. They have shown that one can attain large values of concurrence for inter-emitter distances exceeding resonant wavelength. The graphene SP mediated interaction between two emitters has been recently analyzed [6]. It has been suggested that for this system inter-emitter interactions can be controlled at a subwavelength scale and can be tuned by means of external parameters such as gating voltage.

In the abovementioned schemes, the cooperative behavior of emitters is very sensitive to their spatial position that is related to significant technological challenges. Coupling emitters to zero-index structures [8-12] may enable observation of quantum cooperative effects without necessity of precision positioning of emitters since light experiences no phase advance when propagating in these structures. This observation can be confirmed by the following heuristic argument. In the homogeneous environment, the wavelength corresponding to the resonant frequency defines the spatial scale at which the interaction between emitters is significant. The free space dispersion relation $k = \omega/c$ suggests that increasing the free space wavelength $\lambda$ (or, equivalently, decreasing absolute value of the wave vector $k = 2\pi/\lambda$) inevitably results in decreasing corresponding frequency $\omega$. Here $c$ denotes speed of light in the vacuum. Consider that emitters are embedded in a homogenous material with refractive index $n$. Then, in principle, the smaller values of the refractive index indicate that quantum cooperative behavior will be observed at larger inter-emitter distances, compared to the case when emitters are embedded in the environment with higher refractive index. In this regard one may ask if it is possible to have an environment supporting an electromagnetic wave such that its frequency lies within an optical or telecom band while the wavelength is extremely extended. Such a situation comes up when considering metamaterials with the real part of the effective index close to zero.

Zero-index behavior has been experimentally demonstrated in different systems such as free-standing metallodielectric fishnet nanostructures [12], subwavelength silver and silicon nitride nanolamellae structures [9], and in a metamaterial made of purely dielectric high-index rods [8]. Recently, it has been experimentally demonstrated that rectangular waveguide with metal cladding and dielectric core exhibits effective zero-index behavior at the frequency cutoff [10]. Geometry of the waveguide yields a dispersion relation such that at the frequency cutoff the propagation constant assumes zero value that implies infinite effective wavelength [13, 14]. Moreover, by varying geometrical parameters of the waveguide one can adjust the cutoff frequency such that it lies within optical or telecom band.

To test the heuristic observations presented above, we study cooperative behavior of quantum emitters embedded in a dielectric core of a nanoscale zero-index waveguide (see Fig. 1). We employ rigorous dyadic Green's function-based macroscopic quantum electrodynamics techniques which are well suited to describe quantum emitters coupled to a dispersive and lossy environment [15, 16]. Using finite difference time domain (FDTD) method we calculate collective parameters describing cooperative behavior of quantum emitters coupled to a nanoscale waveguide, collective decay parameter and dipole-dipole coupling [16]. We explore how immersing quantum dipole emitters in a zero-index medium affects superradiance and entanglement generation between quantum emitters.

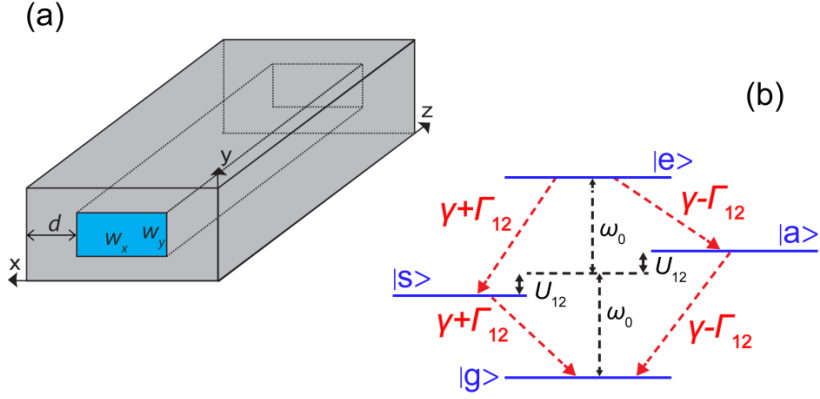

**Figure 1.** (a) A rectangular waveguide with dielectric core and silver cladding. (b) Collective states of two identical atoms with identical decay rates $\gamma$. The energies of symmetric and antisymmetric states ($|s\rangle$ and $|a\rangle$, respectively) are shifted by dipole-dipole coupling $U_{12}$.

We consider a system of $M$ non-identical non-overlapping dipole quantum emitters embedded in a waveguide core and interacting with each other via an electromagnetic field. We model each emitter as a two-level system, with ground state $|g_i\rangle$ and excited state $|e_i\rangle$, where index $i$ specifies the emitter. The transition frequency between the two levels of each emitter is denoted as $\omega_i$, the dipole moments coupling the two levels is denoted as $\vec{d}_i$, and the radial vector indicating the position of the $i^{\text{th}}$ emitter is denoted as $\vec{r}_i$. It is assumed that the emitters are coupled to the modes supported by the waveguide which is initially in the vacuum state. In the weak interaction limit, by applying the Born-Markov and rotating wave approximations, one can derive a master equation obeyed by the reduced density matrix $\rho$. In the Schrodinger picture, the resulting master equation can be written as follows [16]:

$$\frac{\partial}{\partial t}\rho = \frac{1}{i\hbar}[\tilde{H}_0,\rho] - \frac{1}{2}\sum_{i,j=1}^{M}\Gamma_{ij}(\rho S_i^+ S_j^- + S_i^+ S_j^- \rho - 2S_j^- \rho S_i^+), \quad (1)$$

where

$$\tilde{H}_0 = \sum_{i=1}^{M}\hbar(\omega_i + \varepsilon_i)S_i^z + \sum_{i\neq j}^{M} U_{ij} S_i^+ S_j^-. \quad (2)$$

Here $S_i^- = |g_i\rangle\langle e_i|$ and $S_i^+ = |e_i\rangle\langle g_i|$ are the pseudospin operators, $S_i^z$ is defined as $S_i^z = (|e_i\rangle\langle e_i| - |g_i\rangle\langle g_i|)/2$, $\varepsilon_i$ is the reservoir induced frequency shift which we hereafter consider to be absorbed in the frequency $\omega_i$, and parameters $\Gamma_{ij}$ and $U_{ij}$ describe mutual interaction between the emitters. The collective decay rate $\Gamma_{ij}(\vec{r}_i,\vec{r}_j)$ quantifies the variation of the decay rate located at the position $\vec{r}_i$ due to the presence of the second emitter positioned at $\vec{r}_j$. The parameter $U_{ij}$ appears as the proportionality coefficient in the dipole-dipole interaction term and describes coherent coupling of the quantum emitters through the vacuum field. The parameters $\Gamma_{ij}$ and $U_{ij}$ can be expressed in terms of dyadic Green's function of the environment $\ddot{G}(\vec{r}_i,\vec{r}_j,\omega)$ [16, 17]:

$$\Gamma_{ij}(\vec{r}_i,\vec{r}_j) = \frac{2}{\hbar\varepsilon_0}\frac{\omega_j^2}{c^2}\vec{d}_i \operatorname{Im}\ddot{G}(\vec{r}_i,\vec{r}_j,\omega_j)\vec{d}_j^*, \quad (3)$$

$$U_{ij}(\vec{r}_i,\vec{r}_j) = \frac{1}{\hbar\varepsilon_0}\frac{\omega_j^2}{c^2}\vec{d}_i \operatorname{Re}\ddot{G}(r_i,r_j,\omega_j)\vec{d}_j^* + \int_0^\infty dx \frac{x^2}{c^2}\vec{d}_i \operatorname{Re}\ddot{G}(r_i,r_j,ix)\vec{d}_j^* \frac{\omega_j^2}{x^2+\omega_j^2}\}. \quad (4)$$

When $i = j$, $\Gamma_{ij}$ reduces to the single emitter decay rate $\Gamma_{ii} = \gamma$. Taking into account that Green's function obeys reciprocity relation $\vec{\vec{G}}(\vec{r}, \vec{r}', \omega) = \vec{\vec{G}}^T(\vec{r}', \vec{r}, \omega)$ (the superscript $T$ denotes matrix transposition), we conclude that $\Gamma_{12} = \Gamma_{21}$ and $U_{12} = U_{21}$.

In what follows we focus on the case of two identical emitters: $\omega_1 = \omega_2$ and $\vec{d}_1 = \vec{d}_2$. In this case, it is convenient to work in Dicke basis, $|g\rangle = |g_1\rangle|g_2\rangle$, $|a\rangle = (|e_1\rangle|g_2\rangle - |g_1\rangle|e_2\rangle)/\sqrt{2}$, $|s\rangle = (|e_1\rangle|g_2\rangle + |g_1\rangle|e_2\rangle)/\sqrt{2}$, $|e\rangle = |e_1\rangle|e_2\rangle$, which diagonalizes the effective Hamiltonian (2). As one can see from schematics of Fig 1(b), the ground state $|g\rangle$ and excited state $|e\rangle$ are not affected by dipole-dipole interaction while antisymmetric and symmetric states, $|a\rangle$ and $|s\rangle$, correspondingly, are shifted from their unperturbed energies by $\pm U_{12}$. There are two uncorrelated decay channels from the excited state: (i) the "superradiant" channel $|e\rangle \to |s\rangle \to |g\rangle$ with decay rate $\gamma + \Gamma_{12}$, (ii) the "subradiant" channel $|e\rangle \to |a\rangle \to |g\rangle$ with decay rate $\gamma - \Gamma_{12}$.

We use FDTD to calculate the dyadic Green's function of the rectangular waveguide with dielectric core and silver cladding, immersed in the free space. The cross section of the dielectric cladding is taken as 80×240nm, and the thickness of the silver is taken as 250nm. The silver is modeled by using the Palik data [18], and the refractive index of the dielectric core is taken as $n_d = 1.49$. We assume that the dipole moments of the emitters are unidirectional and perpendicular to the line connecting the emitters (in the adopted notations, $y$ oriented) and the emitters are embedded in the center of the infinitely long waveguide. Note that for the frequencies of interest, that is, for the frequencies around the cutoff frequency and slightly above it, the waveguide supports no modes to which $x$ and $z$ polarized dipoles could couple [19]. We compare the collective decay parameter for the emitters embedded in the dielectric core of the waveguide with those in the case when emitters are embedded in a *homogeneous* dielectric of the same refractive index $(n_d = 1.49)$. In this case the collective decay parameter and dipole-dipole coupling can be written analytically, which, for the given configuration of emitter dipole moments takes the following form:

$$\Gamma_{12} = \frac{3}{2} n_d \gamma_0 \left( \frac{\sin(k_B r_{12})}{k_B r_{12}} + \frac{\cos(k_B r_{12})}{(k_B r_{12})^2} - \frac{\sin(k_B r_{12})}{(k_B r_{12})^3} \right), \qquad (5)$$

$$U_{12} = \frac{3}{4} n_d \gamma_0 \left( \frac{\cos(k_B r_{12})}{k_B r_{12}} - \frac{\sin(k_B r_{12})}{(k_B r_{12})^2} - \frac{\cos(k_B r_{12})}{(k_B r_{12})^3} \right), \qquad (6)$$

where $k_B = (\omega/c) n_d$ is the wave vector in the homogeneous dielectric, and $\gamma_0$ is the decay rate in free space. Figure 2(a) displays collective decay rate (5) and dipole-dipole coupling (6) as a function of normalized inter-emitter distance $r_{12}/\lambda_0$, where $r_{12}$ is inter-emitter distance, and $\lambda_0$ is a free space resonant wavelength. As one can see, to observe considerable collective effects in the homogeneous environment with $n_d = 1.49$, one should place emitters very close to each other since collective parameters are negligible for inter-emitter distances larger than a half resonant wavelength $\lambda_0/2$.

To understand collective behavior of emitters embedded in the dielectric core of a rectangular waveguide [see Fig. 1(a)], we first plot single emitter spontaneous emission decay rate enhancement as a function of frequency [see Fig. 2(b)]. Spontaneous emission enhancement peaks at the frequency corresponding to the cutoff frequency of the waveguide. In what follows we analyze collective behavior of two quantum emitters when the resonant frequency of emitters is (i) $\lambda_0 = 1086$ nm which is slightly below the cutoff frequency, (ii) $\lambda_0 = 1071$ nm which is slightly above the cutoff frequency, and (iii) $\lambda_0 = 999$ nm that lies further away from the cutoff frequency. The mentioned resonant wavelengths are marked by vertical lines in Fig. 2(b). Figures 2(c) and 2(d) display collective decay rate and dipole-dipole coupling for emitters embedded in the center of the waveguide core as a function of normalized inter emitter spacing. As one can see, for resonant wavelengths close to the cutoff wavelength collective parameters are non-oscillatory functions of inter-emitter spacing since for considered frequencies the effective wavelength is very large. The oscillations in collective parameters come up when resonant frequencies of emitters are considerably above the cutoff frequency (see the curve corresponding to $\lambda_0 = 999$ nm). Finally, we note that in Figs. 2(a), 2(c), and 2(d) the collective parameters are normalized to the single emitter decay rates in the environments in which emitters are embedded. For example, in case of the emitter embedded in the homogeneous environment with refractive index $n_d$, $\gamma = n_d \gamma_0$.

Figure 2(c) shows that for the resonant wavelength of the emitter $\lambda_0 = 1086$ nm, which is below the cutoff frequency of the waveguide, the collective decay rate $\Gamma_{12}(r_{12})$ adopts non-negligible positive values when $r_{12} < 2.5\lambda_0$. This implies that when studying spontaneous emission dynamics of two excited quantum emitters located within two and a half free space resonant wavelengths from each other, the decay rate of the channel $|e\rangle \to |s\rangle \to |g\rangle$, $\gamma + \Gamma_{12}$, is going to exceed the decay rate of an isolated emitter $\gamma$, and the system is going to exhibit superradiant behavior [see Fig 1(b)]. To quantify the radiation power enhancement due to collective effects, we calculate the ratio of the radiation power of two interacting emitters over the radiation power of two non-interacting emitters embedded in the same environment. To this end, we take into account that the energy of the two-emitter system is given as $E = 2\hbar\omega_0\rho_{ee} + (\hbar\omega_0 - \hbar U_{12})\rho_{ss} + (\hbar\omega_0 + \hbar U_{12})\rho_{aa}$, where $\rho_{ee} = \langle e|\rho|e\rangle$, $\rho_{ss} = \langle s|\rho|s\rangle$, and $\rho_{aa} = \langle a|\rho|a\rangle$ give the probability of the two-atom system to be in the excited, symmetric, and antisymmetric state, correspondingly. Hence, total radiation power from the two-atom system can be calculated as $I(t) = -dE/dt$. Taking into account equations of motion for the density matrix elements [1] and neglecting the terms proportional to dipole-dipole coupling due to $U_{12} \ll \omega_0$, we arrive at the following expression for the average number of photons emitted per unit time $I(t)/(\hbar\omega) = 2\gamma\rho_{ee} + (\gamma + \Gamma_{12})\rho_{ss} + (\gamma - \Gamma_{12})\rho_{aa}$. Figure 3 shows radiation power enhancement due to collective effects $I/I_0$ as a function of normalized distance $r_{12}/\lambda_0$ and normalized time $t\gamma$ for different situations. We assume that at the initial point of time $t = 0$ both emitters are excited, and one of the emitters is placed at $r_{12} = 0$. Figure 3(a) corresponds to the power enhancement in the case of two emitters embedded in a homogeneous dielectric $n_d = 1.49$. In this case, for observation of superradiance, inter-emitter distances should be such that $r_{12} < \lambda_0/5$. Situation drastically changes when considering radiation power enhancement for two emitters embedded in the nanoscale waveguide [Figs. 3(b) and 3(c)]. In this case, when the resonant frequency of emitters lies slightly below the cutoff frequency of the waveguide [Fig. 3(b)], one observes superradiant behavior when emitters are located anywhere within resonant wavelength from each other. Due to large effective wavelength of the supported mode, the constraint on precise positioning of emitters is relaxed. When the resonant frequency of emitters is considerably above the cutoff frequency [Fig. 3(c)], one can still observe superradiance at large inter-emitter distances, however, in this case precision positioning of emitters is necessary. Note that in Fig. 3(c) the peaks in the power enhancement correspond to the peak and dip in collective decay rate $\Gamma_{12}$ [Fig. 2(c)]. This is due to the fact that when $\Gamma_{12} < 0$ previously subradiant channel $|e\rangle \to |a\rangle \to |g\rangle$ becomes superradiant resulting in the spontaneous emission decay rate enhancement. Figure 3(d) depicts the power enhancement at $t = \gamma/2$ as a function of normalized inter-emitter distance for emitters embedded in homogeneous dielectric with $n_d = 1.49$ as well as for emitters embedded in the waveguide core. In the former case, each curve corresponds to a different resonant frequency of emitters. Interestingly, the optimal frequency for observation of superradiant behavior is always smaller than the cutoff frequency. This observation has been confirmed for different waveguide dimensions.

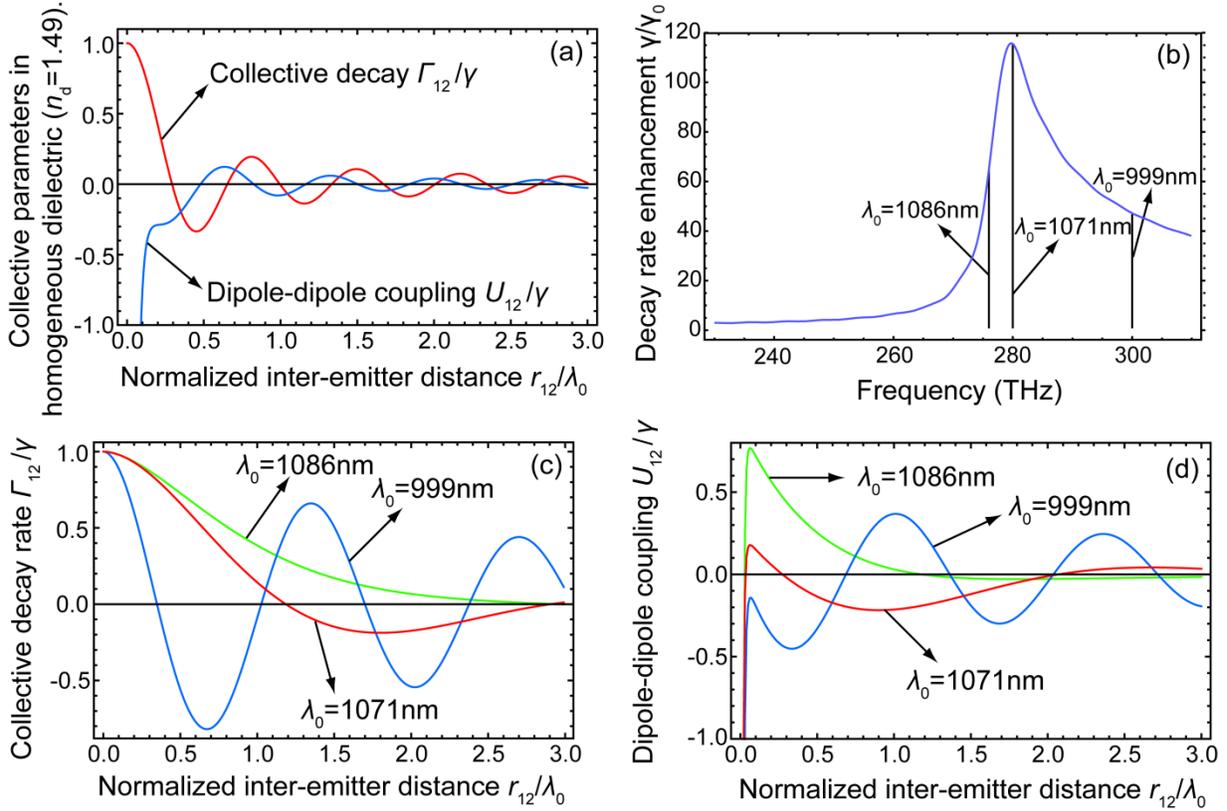

**Figure 2.** (a) Collective decay rate $\Gamma_{12}/\gamma$ and dipole-dipole coupling $U_{12}/\gamma$ as a function of normalized inter-emitter distance for quantum emitters in a homogeneous dielectric with refractive index $n_d = 1.49$. (b) Decay rate enhancement $\gamma/\gamma$ as a function of frequency for a quantum emitter embedded in the waveguide core. Figures 2(c) and 2(d) represent collective decay rate $\Gamma_{12}/\gamma$ and dipole-dipole coupling $U_{12}/\gamma$, correspondingly, for quantum emitters embedded in the waveguide core as a function of normalized inter-emitter distance, for different values of resonant frequencies.

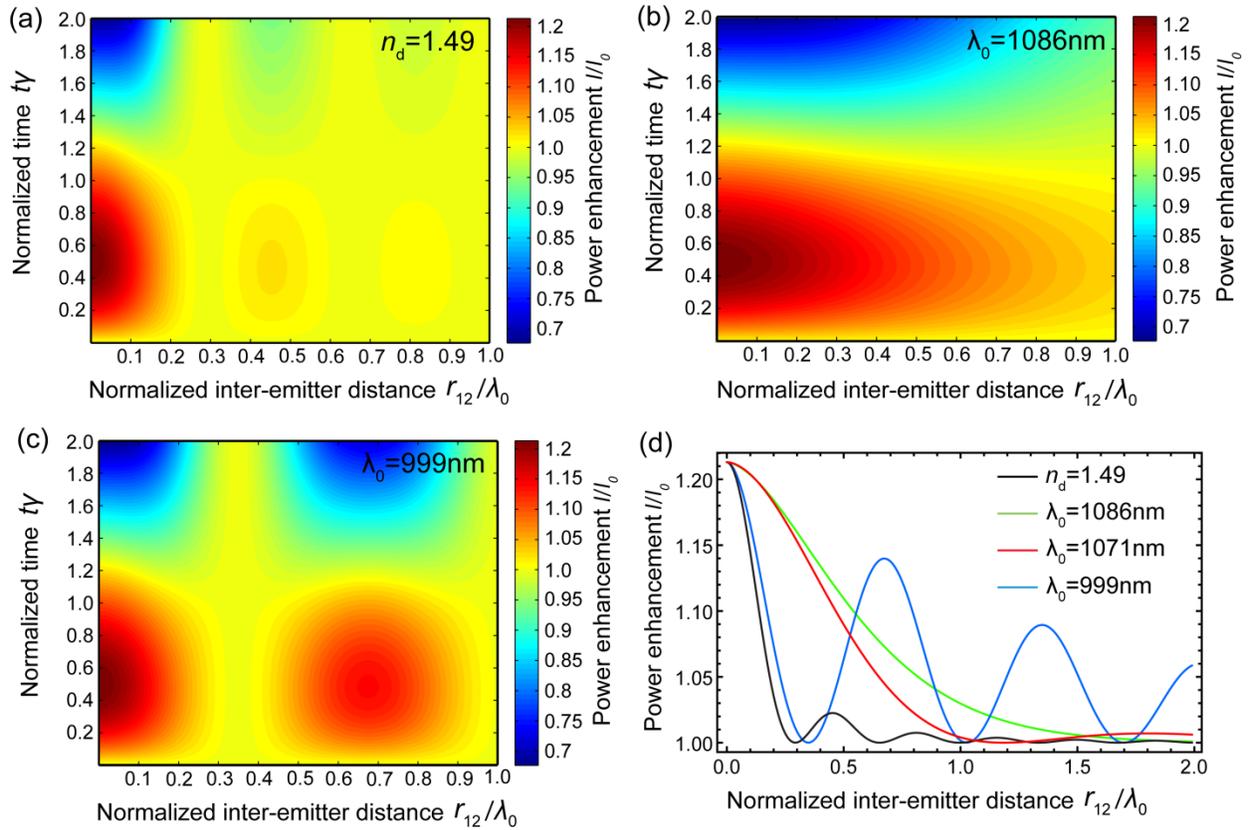

**Figure 3.** Radiation power enhancement due to collective effects as a function normalized inter-emitter distance $r_{12}/\lambda_0$ and of normalized time $\gamma t$. One of emitters is located at $r_{12}=0$. (a) Emitters in free space. (b) Emitters in the waveguide core: the resonant frequency of emitters is slightly below the cutoff frequency ($\lambda_0 = 1086$ nm). (c) Two dipole emitters in the waveguide core: the resonant frequency of emitters is above the cutoff frequency ($\lambda_0 = 999$ nm). (d) Power enhancement at $t=\gamma/2$ as a function of normalized inter-emitter distance for emitters embedded in homogeneous dielectric with $n_d = 1.49$ as well as for emitters embedded in the waveguide core. In the former case, each curve corresponds to a different resonant frequency of emitters.

The superradiant behavior of quantum emitters in a nanoscale channel has been previously discussed [20]. In this work the collective decay parameter of emitters embedded in the waveguide (3) has been approximated by Eq. (5) in which the wave vector $k_B$ has been replaced by the corresponding propagation constant in the waveguide. However, this approach is not valid since, for dipole emitters coupled to a plasmonic waveguide, the parameter $\Gamma_{12}$, as a function of inter-emitter spacing, can crudely be described as an exponentially damped cosine [3], which cannot be approximated by Eq. (5).

It is known that spontaneous emission of two quantum emitters coupled to a common reservoir may result in entanglement generation between the emitters, even if emitters were initially prepared in an unentangled state [21]. In what follows we discuss how coupling emitters to a zero-inex waveguide affects entanglement generation dynamics between quantum emitters. As a measure of entanglement, we use concurrence $C$ introduced by Wooters [22]. The concurrence varies from 0 to 1. For unentangled emitters $C=0$ while for maximally entangled emitters $C=1$. In the context of the superradiance we have discussed a situation when at $t=0$ both emitters are excited. However, exciting two emitters initially is very inefficient for entanglement generation. In what follows we assume that initially only one of the emitters is excited: $\rho_{ee}(0) = 0$, $\rho_{ss}(0) = \rho_{aa}(0) = \rho_{sa}(0) = \rho_{as}(0) = 1/2$. For given initial conditions, concurrence takes the following form:

$$C(t) = \frac{1}{2}\sqrt{[e^{-(\gamma+\Gamma_{12})t} - e^{-(\gamma-\Gamma_{12})t}]^2 + 4e^{-2\gamma t}\sin^2(2U_{12}t)}. \tag{7}$$

As one can see from Eq. (7), $C(0)=0$. For $t > 0$, $C(t)$ becomes positive that means that the emitters become entangled with the degree of entanglement given by (7). The degree of entanglement exponentially goes to zero as $t \to \infty$. Equation (7) also shows that for short enough times, $t < 1/(2\gamma)$, and strong enough dipole-dipole interactions, $U_{12} \gg \gamma$, one may observe oscillations in the concurrence with the oscillation frequency equal to the energy separation between symmetric and antisymmetric states $2U_{12}$ [see Fig. 1(b)]. For large enough times, $t > 1/(2\gamma)$, the only surviving term would be a decaying term with the smallest exponent. Depending on the sign of the collective decay rate $\Gamma_{12}$, the concurrence will either asymptotically approach the population of the antisymmetric state (for $\Gamma_{12} > 0$) or the population of the symmetric state (for $\Gamma_{12} < 0$).

In Figs. 4(a)-4(c) we plot the concurrence as a function of normalized inter-emitter distance $r_{12}/\lambda_0$ and normalized time $\gamma t$, when emitters are coupled to different environments. Fig. 4(a) corresponds to the case when emitters are embedded in a homogeneous dielectric with refractive index $n_d = 1.49$. Figs. 4(b) and 4(c) depict the cases when emitters are embedded in the waveguide core. Figure 4(b) assumes that the resonant frequency of emitters is slightly below the cutoff frequency of the waveguide ($\lambda_0 = 1086$ nm) while Fig. 4(c) corresponds to the case when the resonant frequency of the emitter is above the cutoff frequency ($\lambda_0 = 999$ nm) [see Fig. 2(b)]. In these plots it is assumed that one of the emitters is placed at $r_{12} = 0$. To avoid the divergence of the dipole-dipole coupling at $r_{12} \to 0$, we place the second emitter at inter-emitter distances $r_{12} \geq 0.05\lambda_0$. To facilitate the comparison of Figs. 4(a)-4(c) we fixed the same variation range for the concurrence that led to the saturation in the color scale in Fig. 4(a). Unlike the case when emitters are coupled to the waveguide, for emitters embedded in the homogeneous dielectric, one observes oscillations in the concurrence for smaller time and small inter-emitter separations. On the other hand, when emitters are embedded in the waveguide core one can entangle emitters at larger inter emitter distances. Comparison of Figs. 4(b) and 4(c) shows when the resonant frequency of emitters is around the cutoff frequency of the waveguide, for inter-emitter separations $r_{12} < 0.55\lambda_0$, the concurrence stays longer in the system as compared to the case when resonant frequency of emitters is well above the cutoff frequency. This is due to large absolute values attained by the collective decay rate $\Gamma_{12}$ [Fig. 2(c)]. Figure 4(d) shows concurrence as a function of normalized time $\gamma t$ for a given inter-emitter spacing: $r_{12}/\lambda_0 = 0.067$. Nevertheless emitters in the homogeneous dielectric environment attain higher values of concurrence for shorter times, the concurrence decays faster as compared to the case when emitters are coupled to the waveguide, especially as compared to the case when the waveguide is operating in the zero-index regime ($\lambda_0 = 1086$ nm).

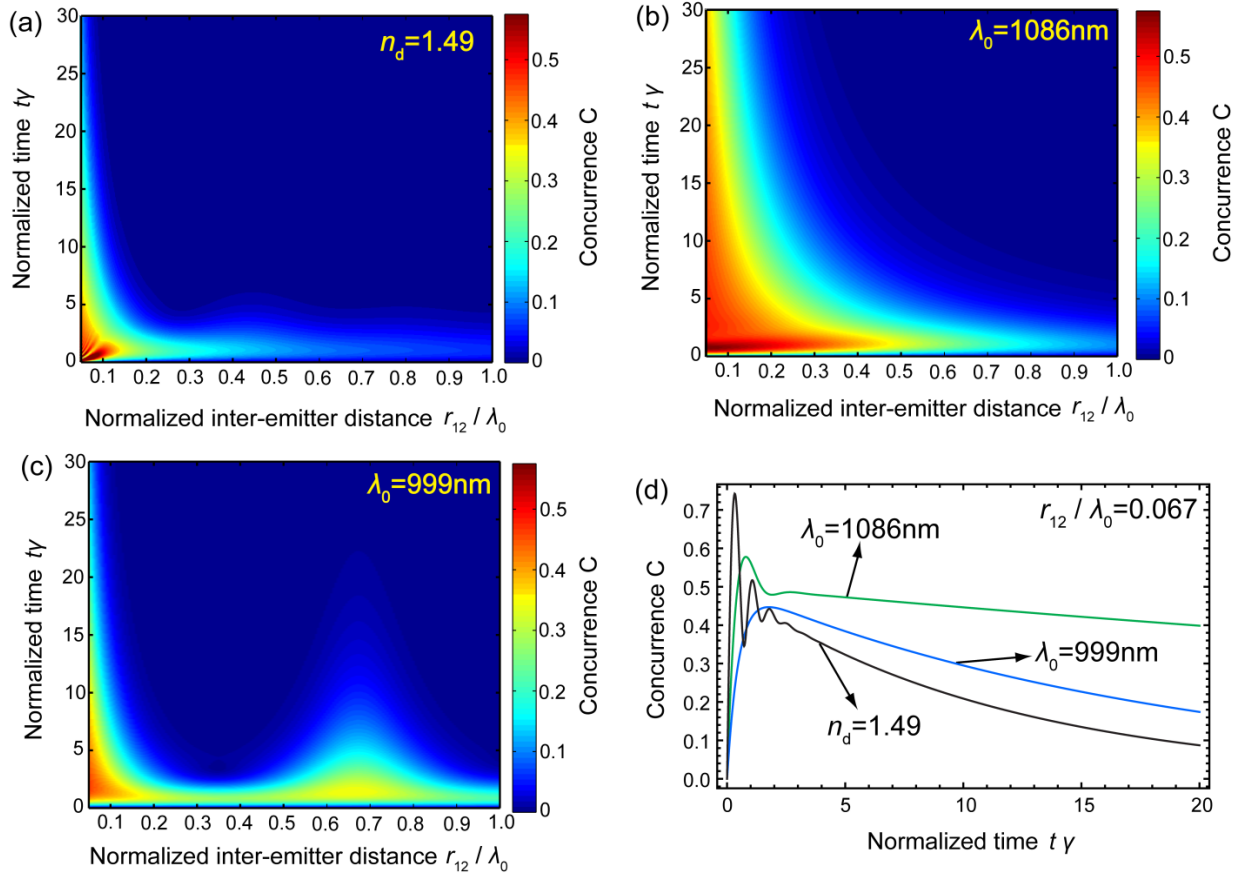

**Figure 4.** Figs. 2(a)-2(c) display concurrence as a function normalized inter-emitter distance $r_{12}/\lambda_0$ and normalized time $\gamma t$. (a) Two dipole emitters in free space. (b) Two dipole emitters in the waveguide core: the resonant frequency of emitters is slightly below the cutoff frequency ($\lambda_0 = 1086$ nm). (c) Two dipole emitters in the waveguide core: the resonant frequency of emitters is above the cutoff frequency ($\lambda_0 = 999$ nm). (d) Concurrence as a function of normalized time $\gamma t$ for normalized inter-emitter spacing $r_{12}/\lambda_0 = 0.067$.

In conclusion, coupling quantum emitters to a zero-index waveguide significantly alters characteristics of cooperative behavior of quantum emitters giving rise to interesting spatial and temporal effects. Due to large effective wavelength of the light supported by a zero-index waveguide, one can significantly extend the spatial volume for which the superradiance effect is observable thus relaxing the constraint of precision positioning of emitters. This may considerably facilitate experimental observation of superradiance in the waveguide when large number of emitters is involved. When studying entanglement generation between two quantum emitters coupled to a zero-index waveguide, we have shown that it is possible to obtain long-term entanglement generation in the system with relatively large values of concurrence. Importantly, also in this case, one doesn't need to be concerned with precision positioning of emitters.

AUTHOR INFORMATION

**Corresponding Author**

*E-mail: sokhoyan@caltech.edu

ACKNOWLEDGMENT

The authors gratefully acknowledge support from the Air Force Office of Scientific Research Quantum Metaphotonic MURI program under grant FA9550-12-1-0488.